\documentclass[english,final,proceeding]{jcaa}
\usepackage{array}
\usepackage{tabu}
\usepackage{subfig}
\title{Limitations of Source-Filter Coupling In Phonation}
\author{Debasish Ray Mohapatra \thanks{d.mohapatra@alumni.ubc.ca}}
\author{Sidney Fels}%\thanks{ssfels@ece.ubc.ca}}
\affil[1]{The University of British Columbia, Vancouver, BC, V6T 1Z4}

\begin{document}

\twocolumn[%%

\maketitle

]%%

\saythanks

\section{Introduction}

As per the traditional source-filter theory, every acoustic speech synthesizer requires a voice source model to produce acoustic energy and a filter which could modulate that energy to produce speech like sound. These models could be categorized into three sections: parametric glottal flow models, kinematic vocal fold models and self-oscillating bio-mechanical vocal fold models.  The parametric glottal flow model assumes that the voice source and vocal tract \textit{filter} are linearly separable. But the current research in speech science precisely illustrates the impact of acoustic loading on the dynamic behaviour of the vocal fold vibration as well as the variation in the glottal flow pulses' shape. Both the kinematic and self-oscillating vocal fold models consider the source-filter interaction. 

This study outlines the source-filter theory; elucidates various low-dimensional lumped-mass models of the acoustic source and computational models of the vocal tract as articulation. To understand the limitations of source-filter interactions which are associated with each of these models, we considered their mechanical design, acoustic and physiological properties and aerodynamic simulation.

\section{Nonlinear Source-Filter Interaction}
The acoustic interaction between the vocal cord and vocal tract is a growing interest in the study of articulatory speech production. The source-filter interaction refers to the properties of the vocal tract model which affect the self-oscillating characteristics of the acoustic source. These properties play a significant role in the designing of an articulatory speech synthesizer. In literature the source filter interaction has been demonstrated by considering the following effects: skewness in glottal flow wave, truncation, dispersion and superposition.

In retrospect, the classic linear source-filter theory \cite{gunnar1960acoustic} assumes human speech generation as a two-stage independent process. First, the glottal source produces air pulses of multiple fundamental frequencies $F_{0}$ which traverses through the vocal tract \textit{(filter)}. The vocal tract acts like an acoustic modulator and it resonates only at formant frequencies $F_{1}$ to produce a time-varying glottal flow as output. This process has been demonstrated in Fig. 1. The linearly separable source-filter models could be represented mathematically as the convolution of the source and filter function in time domain or multiplication in frequency domain.

Though the linearity assumption for source-filter coupling is useful to build an over-simplified model, physiological systems are generally non-linear. And the linear coupling is only suitable where the fundamental frequencies of the source do not cross over the formant frequencies like in male speech voice. But for female or child speech voice and even while singing, it has been observed that the fundamental frequencies are in the proximity of the formant frequencies. And during that case, there is a high degree of variation in the vocal tract impedance which may cause intense interaction between the source and filter models. That could possibly lead to bifurcations in the dynamics of vocal fold vibration, sudden $F_{0}$ jumps and variation in the source energy \cite{titze2008nonlinear}. Hence, to produce the actual speech like sound we should consider the nonlinear and time varying characteristics of source-filter coupling.

\begin{figure}
	\centering
	\includegraphics[width=8cm, height=10cm, keepaspectratio,]{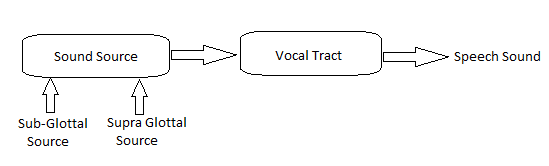}
	\caption{Flow diagram showing the Source-Filter process}
\end{figure}

\section{Speech Synthesizer Models}

There is a multitude of source-filter models in the literature which are successfully implemented. Most of these self-oscillating source models could be organized in two categories: lumped-element and continuum mass models. Though continuum models provide a better representation of the vocal fold, they are complex and computationally expensive. So for simplification, we analyzed only the lumped-element models of the vocal fold. Despite their simplicity, these models were shown to be able to generate many characteristics of an actual vocal fold oscillation. They represent the vocal fold as point masses which are connected to a rigid wall through springs \cite{birkholz2011survey}. Fig. 2 demonstrates the lumped-element models. To analyze the effect of acoustic loading, various computational models of the vocal tract are coupled with the acoustic source models.  %Table 1 shows the models we analyzed for source-filter coupling. 
% \begin{table}
% \caption{Source-Filter Models}
% \label{table_14}
% \begin{tabular}{ccc}
% \hline
% Source Model & Filter Model \\
% \hline
% one-mass model & faeta \\
% \hline
% two-mass model & faeta \\
% \hline
% body-cover model & faeta \\
% \hline
% \end{tabular}
% \end{table}

\subsection{Source Models}
The one-mass model of the vocal fold was designed with a single mass-spring oscillator, driven by airflow from lungs. The point mass is assigned with a fixed weight which could emulate the vocal fold. And the spring system helps the point mass to oscillate to create air pulses with a particular glottal frequency \cite{flanagan1968self}. Although the model can simulate acceptable voiced sound for an inductive acoustic load, it fails to sustain the self-oscillatory behaviour of the source for a capacitive load of the vocal tract, i.e. when the fundamental frequencies are just above the formants \cite{birkholz2011survey}. Because the one-mass model has only one degree of freedom, it could not produce the phase difference between the upper and lower vocal cord edges during oscillation.

Unlike the one-mass model, the two mass model of vocal fold \cite{ishizaka1972synthesis} has successfully demonstrated the self-oscillating characteristics of the vocal fold. And the oscillation sustains for both the inductive and capacitive load of the vocal tract. Two mass model uses two mass elements per vocal cord and provides the necessary degree-of-freedom to introduce the vertical phase difference in vocal fold edges during oscillation. However, the main disadvantage of both of these models is that their tissue discretization in a coronal plane does not capture the layered structure of the vocal folds. And there is no immediate correlation between the spring stiffness and the effects of muscle contractions. From speech synthesis point of view, the most significant factor in the two-mass model is the characterization of its performance when its coupled with a transmission line model of the vocal tract.To investigate the performance of a two-mass model, Ishizaka et al.  measured the onset frequency of the jumps for a large acoustic load on the model and compared the result with the same acoustic load in human voicing. For the two-mass model, the jump in the fundamental frequency happens at the first resonant frequency of the tube. Whereas in human voicing onset frequency of the jumps is higher than the first formant frequency \cite{ishizaka1972synthesis}.

The limitation of layered structured representation in two-mass type models motivated to design the body-cover structure of vocal folds. Mostly its a three-mass model that adds a ``body" mass lateral to the two cover masses. Though the structural representation of the body-cover model is different, this model still preserves the self-oscillatory principle of the two-mass model. By controlling the body stiffness constant, the body-cover model of vocal fold could be reduced to a two-mass model.

\begin{figure}%
    \centering
    \subfloat{{\includegraphics[width=3cm, height=2.5cm, keepaspectratio]{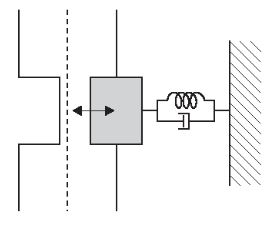} }}%
    \qquad
    \subfloat{{\includegraphics[width=3cm, height=2.5cm, keepaspectratio]{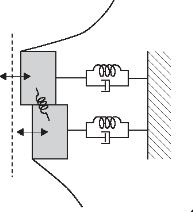} }}%
    \qquad
    \subfloat{{\includegraphics[width=3cm, height=2.5cm, keepaspectratio]{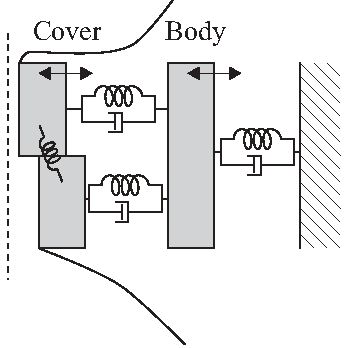} }}%
    \caption{Lumpled-Element models of vocal fold. Image by Peter Birkholz \cite{birkholz2011survey}}%
    \label{fig:example}%
\end{figure}

\subsection{Filter Models}
The vocal tract acoustic load has both positive and negative types of damping terms. The negative damping which helps in vocal fold oscillation, is provided by the vocal tract inertance. There are numerous articulatory models exist in the literature which could produce this favourable condition. The Kelly-Lochbaum model of the vocal tract is traditionally constructed by approximating the cross-sectional area of the vocal tract by cascading multiple cylindrical tube sections \cite{valimaki1994improving}. The resulting tube system then could be interpreted as a digital waveguide or digital wave model of the vocal tract. But the severe pitfalls in this model are: 1) length of each of the cylindrical tube section has to be equal. 2) the junction of two tube section is not smooth. These limitations affect the formant frequencies of the articulatory model which in turn prevent to achieve the exact matching of a given speech spectrum.

\section{Conclusions}

The limitation in source-filter interaction for various models has been discussed. One of the significant challenges in designing a speech synthesizer model is to address the degree of interaction between the vocal fold and vocal tract which varies based on the types of articulatory gestures like singing; breathy voice; male or female voice; high and low pitch voice. So it is much needed to do an accurate measurement of the shape changes in vocal tract during articulation and the precise simulation of vocal fold and vocal tract interaction using a feedback channel. Hence, for a specific change in the vocal tract shape, there could be a notable impact on the vocal fold oscillation and glottal flow which is an input to the resonator. A fruitful direction could be the use of 2D Finite-Difference Time-Domain wave solver and the excitation mechanism while maintaining the stability of the solver when the domain boundaries (i.e., the vocal tract walls) are dynamically modified, as in the case of articulation. 

\section{Acknowledgement}
This work was funded by the Natural Sciences and Engineering Research Council (NSERC) of Canada and Canadian Institutes for Health Research (CIHR).

%\section*{References}

\bibliographystyle{unsrt}
\bibliography{jcaa}
\nocite{*}

\end{document}